\documentclass{eceasst}
\usepackage{color}
\usepackage{amsmath}
\usepackage{graphicx}
\usepackage{url}

\newif\ifdraft
\drafttrue

\newcommand{\comment}[1]{/$\ast$ \textit{#1\/} $\ast$/}
\newcommand{\tannot}[1]{//@#1@//}
\newcommand{\deq}{\mathrel{\stackrel{\scriptscriptstyle\Delta}{=}}} 
\renewcommand{\implies}{\Rightarrow}

\newenvironment{display}{\begin{itemize}\item[]}{\end{itemize}}

\newenvironment{conj}{\begin{array}[t]{@{\ensuremath{\land}\ \ }l@{}}}{\end{array}}
\newenvironment{disj}{\begin{array}[t]{@{\ensuremath{\lor}\ \ }l@{}}}{\end{array}}

\ifdraft
\long\def\ednote#1#2{\begin{quote}\framebox{\begin{minipage}{0.99\linewidth}\footnotesize\color{red} #1: #2\end{minipage}}\end{quote}}
\newcommand{\edmargin}[2]{\marginpar{\raggedright\footnotesize\color{red}#1: #2}}
\else
\long\def\ednote#1#2{}
\newcommand{\edmargin}[2]{}
\fi

\makeatletter
\newcounter{abr@ctr}
\newcommand{\abr@c}{\c@abr@ctr\advance\c@abr@ctr\@ne}

\ifx\documentclass\undefined
\else
  \DeclareSymbolFont{tlaitalics}{\encodingdefault}{cmr}{m}{it}
  \let\itfam\symtlaitalics
\fi

\newcommand{\noTeXmath}{%
\c@abr@ctr=\itfam
\multiply\c@abr@ctr"100\relax
\advance\c@abr@ctr "7061\relax
\mathcode`a=\abr@c\mathcode`b=\abr@c\mathcode`c=\abr@c\mathcode`d=\abr@c
\mathcode`e=\abr@c\mathcode`f=\abr@c\mathcode`g=\abr@c\mathcode`h=\abr@c
\mathcode`i=\abr@c\mathcode`j=\abr@c\mathcode`k=\abr@c\mathcode`l=\abr@c
\mathcode`m=\abr@c\mathcode`n=\abr@c\mathcode`o=\abr@c\mathcode`p=\abr@c
\mathcode`q=\abr@c\mathcode`r=\abr@c\mathcode`s=\abr@c\mathcode`t=\abr@c
\mathcode`u=\abr@c\mathcode`v=\abr@c\mathcode`w=\abr@c\mathcode`x=\abr@c
\mathcode`y=\abr@c\mathcode`z=\abr@c
\c@abr@ctr=\itfam
\multiply\c@abr@ctr"100\relax
\advance\c@abr@ctr "7041\relax
\mathcode`A=\abr@c\mathcode`B=\abr@c\mathcode`C=\abr@c\mathcode`D=\abr@c
\mathcode`E=\abr@c\mathcode`F=\abr@c\mathcode`G=\abr@c\mathcode`H=\abr@c
\mathcode`I=\abr@c\mathcode`J=\abr@c\mathcode`K=\abr@c\mathcode`L=\abr@c
\mathcode`M=\abr@c\mathcode`N=\abr@c\mathcode`O=\abr@c\mathcode`P=\abr@c
\mathcode`Q=\abr@c\mathcode`R=\abr@c\mathcode`S=\abr@c\mathcode`T=\abr@c
\mathcode`U=\abr@c\mathcode`V=\abr@c\mathcode`W=\abr@c\mathcode`X=\abr@c
\mathcode`Y=\abr@c\mathcode`Z=\abr@c}

\makeatother
\noTeXmath

\volume{70}{2014} 
\volumetitle{
Proceedings of the\\
14th International Workshop on\\
Automated Verification of Critical Systems
(AVoCS 2014)}
\volumeshort{
Proc.\ AVoCS 2014}
\guesteds{
Marieke Huisman, Jaco van de Pol}

\title{Analyzing Conflict Freedom For Multi-threaded Programs With Time Annotations%
  \sponsor{This work has been supported by a grant from the Airbus Corporate
    Foundation and complementary funding by R\'egion Lorraine, and this grant
    has funded a post-doctoral contrat for Jingshu Chen.}
}
\short{Analyzing Multi-Threaded Programs With Time Annotations} 

\author{%
  Jingshu Chen\autref{1},
  Marie Duflot\autref{{1,2}},
  Stephan Merz\autref{{1}}
} 

\institute{%
  \autlabel{1}Inria, Villers-l\`es-Nancy, F-54600, France\\
  \autlabel{2}Universit\'e de Lorraine, LORIA, UMR~7503, Vand{\oe}uvre-l\`es-Nancy, F-54500, France
} 

\abstract{Avoiding access conflicts is a major challenge in the design of multi-threaded programs. In the context of real-time systems, the absence of conflicts can be guaranteed by ensuring that no two potentially conflicting accesses are ever scheduled concurrently.

In this paper, we analyze programs that carry time annotations specifying the time for executing each statement. We propose a technique for verifying that a multi-threaded program with time annotations is free of access conflicts. In particular, we generate constraints that reflect the possible schedules for executing the program and the required properties. We then invoke an SMT solver in order to verify that no execution gives rise to concurrent conflicting accesses. Otherwise, we obtain a trace that exhibits the access conflict.
} 

\keywords{multi-threaded program, access conflict, real-time system, time annotation, SMT solving} 

\begin{document}
\maketitle

\section{Introduction}
\label{sec:intro}

Avoiding conflicting accesses to shared resources is a fundamental problem in concurrent programming, and it is particularly crucial in the development of controllers of real-time systems. Whereas the use of locks is the most common solution to this problem, it has well-known drawbacks, such as the run-time overhead associated with acquiring locks and being prone to errors and deadlocks. In real-time systems, the use of locks may be incompatible with the stringent requirements on the predictability of running times. Instead, programmers may rely on temporal conditions that ensure that statements with potentially conflicting accesses to resources are never scheduled concurrently.

In this paper, we assume that program code carries annotations that indicate the execution time allowed for each statement of the program~\cite{TWRealTime04}. Such annotations may for example be derived from a static analysis providing bounds on the execution time of the code on a specific execution platform (such as JOP~\cite{schoeberl:worst-case} for safety-critical Java~\cite{jsr302}, although the focus in this paper is on general principles rather than any specific language). Moreover, we assume that the platform provides mechanisms for ensuring that the actual execution of the program complies with these annotations. Our goal is to ensure that no conflicting accesses occur; such a specification can be expressed by precedence properties between statements of different threads.

We present a technique for verifying whether all finite executions of an annotated multi-threaded program up to a fixed bound satisfy the required precedence properties. Similar to bounded model checking, the key idea is to reduce this verification problem to a constraint solving problem, by encoding the set of possible schedules of the given program that respect the timing annotations, and also the required properties, as formulas in quantifier-free linear integer arithmetic. We then invoke off-the-shelf SMT solvers that efficiently decide the satisfiability of such formulas. In case the properties are violated, the solver generates a (counter-)model that corresponds to an execution violating the property, and this model can be analyzed by the program designer. Since the analysis is completely automatic, and the performance of the SMT solvers scales well, program designers can repeatedly analyze different variants of the program and understand the effect of changing timing parameters.

In this paper, we restrict attention to very simple programs where every thread consists of straight-line code, possibly contained in a single loop (which is unrolled for bounded verification). Such simple program structures are not uncommon in the real-time domain, for example when sensor inputs have to be sampled and processed at regular intervals. An extension to more complex control structures is straightforward by over-approximating the possible executions. For a more precise analysis, our technique could be combined with standard SMT-based program analysis~\cite{filliatre:why3,leino:dafny}.

\paragraph{Outline.}

Section~\ref{sec:moti} presents a motivating example, and Section~\ref{sec:model} describes the model of execution for the programs that we analyze. Section~\ref{sec:cg} represents the core of our paper, where we define how constraints are generated to represent the possible schedules of an annotated program, and its precedence properties. The results of some experiments, providing evidence for the scalability of the approach, are reported in Section~\ref{sec:imple}. Section~\ref{sec:dis} discusses related work, and Section~\ref{sec:conclu} concludes the paper.

\section{A Motivating Example}
\label{sec:moti}

As a toy example, consider the following code snippet where $i$ and $j$ are two global variables. 

\begin{display}
  \begin{tabbing}
    $l_{9,9}:$\hspace*{2mm}\= \hspace{3.5cm}\= $l_{9,9}:$\hspace*{2mm}\=\kill
    \comment{Thread $t_1$}      \>\> \comment{Thread $t_2$}\\
    $l_{1,1}:$ \> \tannot{$1$}        \> $l_{2,1}:$ \> sleep($2$);\\
           \> $i = 2$;            \> $l_{2,2}:$ \> \tannot{$2$}\\
    $l_{1,2}:$ \> \tannot{$2$}        \>        \> $j = i$;\\
           \> $i \mathop{+\!\!=} 2$;\\[.5ex]
    \comment{post-condition:\quad $j == i$}
  \end{tabbing}
\end{display}
 
This program can be viewed as an implementation of the classic \emph{producer-consumer} problem, which is representative for synchronization between threads. The two threads $t_1$ and $t_2$ update the values of the variables $i$ and $j$. It is intended that the values of $i$ and $j$ are equal at the end of the execution of the program, that is, the assignment $l_{2,2}$ in thread $t_2$ should be executed after the statements $l_{1,1}$ and $l_{1,2}$ of thread $t_1$.

The standard means for ensuring thread synchronization is the use of locks. However, the use of locks can be costly and error-prone. For programs written for real-time execution platforms where all threads share a common global time reference, such as Safety-Critical Java~\cite{jsr302}, an alternative is to synchronize threads by scheduling constraints. In the above code, these constraints are indicated by the annotations at each statement, resp.\ by the argument of the sleep statement $l_{2,1}$. For example, the assignment statement $l_{1,1}$ is assumed to be scheduled for execution during exactly one time unit. We require these annotations to be present as an input for our analysis, and we assume that they are enforced by the execution platform. We assume that multi-threaded programs are scheduled on a single processor, subject to an arbitrary, but eager scheduling policy where some thread executes whenever at least one thread is executable. Finally, we do not explicitly consider statements such as input and output that could execute in parallel to the CPU. The question whether the assumed scheduling constraints are feasible is out of the scope of this paper, but upper bounds for the execution of statements on specific processor architectures such as JOP~\cite{schoeberl:worst-case} can be obtained by static analysis.

The annotated program in the above example indeed ensures its post-condition: initially, thread $t_2$ is sleeping, and thread $t_1$ is scheduled to execute $l_{1,1}$ for one time unit. After that, $t_2$ is still sleeping, so $t_1$ must again be scheduled for executing $l_{1,2}$, and only then $l_{2,2}$ can execute. However, if the timing annotation for statement $l_{1,1}$ were changed to $2$, then the two threads would compete for execution after two time units, hence $l_{2,2}$ could be scheduled for execution in between statements $l_{1,1}$ and $l_{1,2}$, leading to a violation of the post-condition.

In the following, we describe an approach for mechanically analyzing schedules of multi-threaded programs with timing annotations, with respect to properties that require temporal orders between program statements, typically ensuring the absence of race conditions for accessing shared variables. We generate constraints that describe the potential schedules, as well as required synchronization properties, and use off-the-shelf SMT solvers for verifying that all schedules respecting the constraints satisfy the properties. Otherwise, the solver generates a model that represents an execution of the program violating the properties.

\section{Execution Model}
\label{sec:model}

The input to our analyzer is a multi-threaded program with timing annotations indicating the time alloted to the execution of (blocks of) statements. We distinguish between ordinary and sleep statements: the latter specify that scheduling of the adjacent ordinary statements must be separated by at least the indicated sleeping time. For simplicity, we assume that each thread consists of a sequence of (ordinary and sleep) statements, possibly enclosed in a loop. Without loss of generality, we assume that no thread contains two consecutive sleep statements: the sequence sleep($m$); sleep($n$) is equivalent to the single sleep statement sleep($m+n$).


We will generate constraints that describe all possible schedules of the program execution, up to a user-defined bound. A thread has four possible states: \emph{executing} (a non-sleep statement), \emph{waiting}, \emph{sleeping}, and \emph{terminated}. Threads are scheduled according to the following constraints:
\begin{itemize}
\item At any given instant, at most one thread is in state \emph{executing}. That thread executes its current statement (or block of statements) without interruption by other threads, for the number of time units indicated by the corresponding timing annotation. After that lapse of time, the scheduler may choose to schedule a different thread for execution.
\item Whenever there is at least one thread that is neither sleeping nor terminated, then some thread is executing.
\item A statement sleep($n$) following an ordinary statement causes the thread to enter the sleeping state as soon as the preceding statement has finished executing, and to remain in sleeping state for $n$ time units. After that lapse of time, the thread moves to state waiting, unless it is immediately scheduled for execution or it has terminated. The scheduling of an initial statement sleep($n$) is analogous, at the beginning of program execution. In particular, any number of threads may be sleeping simultaneously.
\item Statements of every thread are scheduled in program order.
\end{itemize}
We leave relaxations of these constraints as interesting topics for future work. In particular, the execution semantics of modern programming languages on advanced architectures, including multi-core or multi-processor systems, does not adhere to all of the above assumptions.


\section{Constraint Generation}
\label{sec:cg}

We now describe constraints that encode the set of possible schedules for a given program, up to a fixed bound. We first list the variables that we use for representing schedules, then give a formula that represents the execution of an individual statement, and finally define the overall scheduling constraints as well as the formula representing the precedence properties to be verified.

\subsection{Representing Program Schedules}
\label{sec:variables}

Suppose that we are given a program with threads $Td = \{1, \ldots, T\}$, and that we want to represent schedules of length up to $N$ steps. We eliminate loops by unrolling every loop so that every thread $t$ consists of statements $s_{t,1}, \dots, s_{t,n_t}$ executed sequentially. The number of ordinary (i.e., non-sleeping) statements in that sequence should be $N$, unless thread $t$ has less than $N$ such statements to execute even when loops are unrolled. We denote by $D_{t,i}$ be the duration of statement $s_{t,i}$, given as an integer constant that corresponds either to the timing annotation if $s_{t,i}$ is non-sleeping, or to the argument of the sleep statement $s_{t,i}$. Let $NS_t \subseteq \{1, \dots, n_t\}$ denote the set of the corresponding indices for non-sleeping statements of thread $t$.

Our encoding is based on the following variables:
\begin{itemize}
\item $pc_t^{(k)}$, for $t \in Td$ and $k \in \{1, \dots, N+1\}$, represents the ``program counter'' of thread $t$. Its value in $NS_t \cup \{n_t+1\}$ denotes the next non-sleeping statement that thread $t$ will execute at round $k$ of the schedule; the value of $n_t+1$ corresponds to a terminated thread.
\item $Y^{(k)}$ and $X^{(k)}$, for $k \in \{1, \dots, N+1\}$,\footnote{The variables $Y^{(N+1)}$ and $X^{(N+1)}$ could be omitted, but their presence yields more uniform definitions.} indicate the global time at the beginning and the end, respectively, of round $k$ of the schedule. Except in situations where all threads are sleeping or have terminated, we will have $Y^{(k+1)} = X^{(k)}$.
\item $E_{t,i}$, for $t \in Td$ and $i \in \{1, \dots, n_t\}$, denotes the time at which the execution of statement $s_{t,i}$ ends (the starting time of execution is then obtained as $E_{t,i} - D_{t,i}$). Observe that we have only one copy of these variables since each statement is executed at most once. Since the schedule ends after $N$ rounds, only the values of $E_{t,i}$ corresponding to statements that have actually been scheduled, are meaningful.
\end{itemize}

The following formula $init$ fixes some values for variables corresponding to the initial round of the schedule.\footnote{We adopt the convention of writing multi-line conjunctions and disjunctions as lists bulleted with the operation sign, using indentation for indicating precedence~\cite{lamport:howtowrite}.}

\[
  init\ \deq\ 
  \begin{conj}
    \displaystyle\bigwedge_{t \in Td} pc_t^{(1)} = 
       \left\{\begin{array}{ll}
           1 & \text{if } 1 \in NS_t\\
           2 & \text{otherwise}
       \end{array}\right.\\
    Y^{(1)} = 
      \left\{\begin{array}{ll}
          0 & \text{if } 1 \in NS_t \text{ for some } t \in Td\\
          \min \{ D_{t,1} : t \in Td \} & \text{otherwise}
      \end{array}\right.\\
    \displaystyle\bigwedge_{t \in Td : 1 \notin NS_t} \!\!\!\!\!\!\!\! E_{t,1} = D_{t,1}
  \end{conj}
\]
The program counters of each thread are initialized to the first non-sleeping statements. The global time at which the first round starts is $0$, except if the initial statements of all threads are sleep statements, in which case the first round starts at the end of the sleep statement(s) with the shortest duration. Finally, all initial sleep statements end after the sleeping time has elapsed.

\subsection{Modeling Execution of a Non-Sleeping Statement}
\label{sec:execute-constraint}

We now define a formula $exec_{t,i}^{(k)}$ that models execution of the non-sleeping statement $s_{t,i}$ (i.e., for $i \in NS_t$) at round $k$. If statement $s_{t,i}$ is not followed in thread $t$ by a sleep statement, the formula is defined as
\[
  exec_{t,i}^{(k)}\ \deq\ 
  \begin{conj}
    pc_t^{(k)} = i \\
    X^{(k)} = Y^{(k)} + D_{t,i}\ \land\ E_{t,i} = X^{(k)}\\
    pc_t^{(k+1)} = i+1\\ 
    \displaystyle\bigwedge_{t' \in Td \setminus \{t\}} \!\!\!\!\! pc_{t'}^{(k+1)} = pc_{t'}^{(k)}
  \end{conj}
\]
Formula $exec_{t,i}^{(k)}$ requires that statement $s_{t,i}$ be the next statement that thread $t$ should execute at round $k$. Then, round $k$ ends at time $Y^{(k)}+D_{t,i}$, which is also the time at which execution of $s_{t,i}$ ends. The program counter for thread $t$ at the next round moves to the subsequent statement, while the other program counters remain unchanged. The starting time of the subsequent round, i.e.\ the value of $Y^{(k+1)}$, will be determined by the overall scheduling constraint defined in Section~\ref{sec:global-constraint}.

The formula $exec_{t,i}^{(k)}$ is somewhat different if statement $s_{t,i}$ is followed by a sleep statement $s_{t,i+1}$: as described in Section~\ref{sec:model}, the sleeping time begins immediately after statement $s_{t,i}$ has been executed, and the next statement to be executed is the statement following $s_{t,i+1}$. We therefore define in this case
\[
  exec_{t,i}^{(k)}\ \deq\ 
  \begin{conj}
    pc_t^{(k)} = i \\
    X^{(k)} = Y^{(k)} + D_{t,i}\ \land\ E_{t,i} = X^{(k)}\ \land\ E_{t,i+1} = X^{(k)} + D_{t,i+1}\\
    pc_t^{(k+1)} = i+2\\
    \displaystyle\bigwedge_{t' \in Td \setminus \{t\}} \!\!\!\!\! pc_{t'}^{(k+1)} = pc_{t'}^{(k)}
  \end{conj}
\]

\subsection{Overall Scheduling Constraint}
\label{sec:global-constraint}

The overall constraint $sched$ characterizing prefixes of schedules of length $N$ asserts that at every round, some non-sleeping statement is executed, unless all threads have (and remain) terminated. This constraint also defines the starting time $Y^{(k+1)}$ for the next round. The following definitions show the high-level structure and the case of termination.
\[\begin{array}{rcl}
  sched & \deq & \displaystyle\bigwedge_{k=1}^N round^{(k)}\\
  round^{(k)} & \deq & terminated^{(k)} \lor exec\_thread^{(k)}\\
  terminated^{(k)} & \deq & 
    \begin{conj}
      \displaystyle\bigwedge_{t \in Td} pc_t^{(k)} = n_t+1\ \land\ pc_t^{(k+1)} = pc_t^{(k)}\\
      Y^{(k+1)} = X^{(k)}\ \land\ X^{(k+1)} = X^{(k)}
    \end{conj}
\end{array}\]
Formula $sched$ stipulates that the schedule should contain $N$ rounds. The constraint characterizing round $k$ distinguishes between two cases: either all threads are already terminated or some thread will execute at round $k$. Termination means that the program counters of all threads point beyond the last statement; they then remain there, and the beginning and ending times of round $k+1$ are set to $X^{(k)}$, the ending time of round $k$.

When program execution has not terminated, some thread $t$ executes a non-sleeping statement, and we must determine the starting time of round $k+1$. Using the formulas defined in Section~\ref{sec:execute-constraint}, this suggests the definition
\[\begin{array}{rcl}
  exec\_thread^{(k)} & \deq &
  \begin{conj}
    \displaystyle\bigvee_{t \in Td} \bigvee_{i \in NS_t} (exec_{t,i}^{(k)} \land (i=1 \lor E_{t,i-1} \leq Y^{(k)}))\\
    fix\_starting\_time^{(k)}
  \end{conj}
\end{array}\]

For the definition of the starting time $Y^{(k+1)}$ of round $k+1$, there are two cases to consider:
\begin{itemize}
\item If some non-sleeping statement can be executed at time $X^{(k)}$, the ending time of round $k$, then $Y^{(k+1)} = X^{(k)}$. There is a statement to be executed at time $X^{(k)}$ iff some non-terminated thread $t$ is either at the beginning of its program 
or execution of the statement preceding the current statement of thread $t$ ended at time $X^{(k)}$ or before.
\item If no statement is executable at time $X^{(k)}$, i.e.\ if all non-terminated threads are sleeping, then round $k+1$ starts when the first thread(s) awake.
\end{itemize}

In the formal definition, we make use of a macro notation in order to refer to the ending time of the statement preceding the current one of a given thread. Specifically, we write $E_{prev_t^{(k)}} \sim e$ where $\mathop{\sim} \in \{=,<,\leq,\geq,>\}$ is a comparison operator, and $e$ is an arbitrary expression, as a shorthand for the formula
\[
  \bigvee_{i=1}^{n_t} \big( pc_t^{(k)} = i+1 \ \land\ E_{t,i} \sim e \big)
\]
and similarly for $e \sim E_{prev_t^{(k)}}$. With this notation, the intuition given above is concretized by the following formulas.

\[\begin{array}{rcl}
  fix\_starting\_time^{(k)} & \deq &
  \begin{disj}
    some\_executable^{(k)} \land Y^{(k+1)} = X^{(k)}\\
    \lnot some\_executable^{(k)} \land set\_min\_end\_time^{(k)}
  \end{disj}\\
  some\_executable^{(k)} & \deq & \displaystyle\bigvee_{t \in Td}
  \begin{conj}
    pc_t^{(k+1)} \neq n_t+1\\
    pc_t^{(k+1)} = 1 \lor E_{prev_t^{(k+1)}} \leq X^{(k)}
  \end{conj}\\
  set\_min\_end\_time^{(k)} & \deq & \displaystyle\bigvee_{t \in Td}
  \begin{conj}
    pc_t^{(k+1)} \neq n_t+1 \\
    \displaystyle\bigwedge_{t' \in Td \setminus \{t\}} \!\!\!\!\! 
       pc_{t'}^{(k+1)} \neq n_{t'}+1 \implies E_{prev_t^{(k+1)}} \leq E_{prev_{t'}^{(k+1)}}\\
    Y^{(k+1)} = E_{prev_t^{(k+1)}}
  \end{conj}
\end{array}\]

Observe that the formula $sched$ is expressed as a quantifier-free formula of the theory of linear integer arithmetic. Off-the-shelf SMT solvers such as Yices~\cite{dutertre:fast} or Z3 \cite{moura:z3} provide very efficient decision procedures for such formulas, from which we can directly benefit.

\subsection{Verifying Precedence Requirements}
\label{sub:spec}

The properties that we are interested in assert precedence between the execution order of statements of different threads, such as that some statement should be executed before another one, or that it should not be executed in between two other statements. Formally, such properties can be expressed as Boolean combinations $\lambda$ of inequations between the ending times for statements. For the toy example of Section~\ref{sec:moti}, we want to assert that the second statement $l_{1,2}$ of thread $t_1$ ends before the second statement $l_{2,2}$ of thread $t_2$ starts. Since two statements are never executed at the same time, this requirement can be expressed as $E_{1,2} < E_{2,2}$.
However, ending times are meaningful only if the corresponding statements have actually been scheduled, and we therefore actually generate the formula
\[
  (pc_1^{(N+1)} > 2\ \land\ pc_2^{(N+1)} >2)\ \implies\ E_{1,2} < E_{2,2}.
\]

Moreover, in case the statements of interest appear in loops, we actually want to verify that such constraints hold for all pairs of instances of these statements when the loops are unrolled.

In order to verify that the property holds over all possible schedules, we generate the formula $sched \land \lnot\lambda$ and run an SMT solver that checks if that formula is satisfiable. If the answer is UNSAT, then the precedence property $\lambda$ holds over all prefixes of schedules of size (at most) $N$. Otherwise, the model computed by the SMT solver corresponds to a schedule that includes the relevant statements and that violates $\lambda$.

\section{Experiments}
\label{sec:imple}

We now illustrate the approach described in Section~\ref{sec:cg}. We have developed a prototype that generates the constraints corresponding to given programs with timing annotations, as well as the desired precedence properties. In our experiments, we use the SMT solver Yices~\cite{dutertre:fast}.

\subsection{Generating the Constraints for a Toy Program}
\label{sec:example-1}

We will generate the constraints for the toy example considered in Section~\ref{sec:moti}.

\begin{display}
  \begin{tabbing}
    $l_{9,9}:$\hspace*{2mm}\= \hspace{3.5cm}\= $l_{9,9}:$\hspace*{2mm}\=\kill
    \comment{Thread $t_1$}        \>\> \comment{Thread $t_2$}\\
    $l_{1,1}:$ \> \tannot{$1$}       \> $l_{2,1}:$ \> sleep($2$);\\
           \> $i = 2$;               \> $l_{2,2}:$ \> \tannot{$2$}\\
    $l_{1,2}:$ \> \tannot{$2$}        \>           \> $j = i$;\\
           \> $i \mathop{+\!\!=} 2$;\\[.5ex]
    \comment{post-condition:\quad $j == i$}
  \end{tabbing}
\end{display}

Since this program has three non-sleeping statements, we generate the constraints representing the possible schedules of length $N=3$. The initial constraint is
\[
  init\ \deq\ 
  \begin{conj}
    pc_1^{(1)} = 1 \ \land\ pc_2^{(1)} = 2\\
    Y^{(1)} = 0\\
    E_{2,1} = 2
  \end{conj}
\]
This constraint initializes the program counters for the two threads to their first non-sleeping statements. Since the initial statement of thread $1$ is non-sleeping, the first round will start at time $0$. The initial (sleep) statement of thread $2$ ends after $2$ time units.

Next, we define formulas that represent the execution of the three non-sleeping statements at round $k$, for $k=1,2,3$, according to the schema given in Section~\ref{sec:execute-constraint}.

\[\begin{array}{@{}l}
  exec_{1,1}^{(k)}\ \deq\ 
  \begin{conj}
    pc_1^{(k)} = 1\\
    X^{(k)} = Y^{(k)} + 1\ \land\ E_{1,1} = X^{(k)}\\
    pc_1^{(k+1)} = 2\ \land\ pc_2^{(k+1)} = pc_2^{(k)}
  \end{conj}\\
  \ \\
  exec_{1,2}^{(k)}\ \deq\ 
  \begin{conj}
    pc_1^{(k)} = 2\\
    X^{(k)} = Y^{(k)} + 2\ \land\ E_{1,2} = X^{(k)}\\
    pc_1^{(k+1)} = 3\ \land\ pc_2^{(k+1)} = pc_2^{(k)}
  \end{conj}\\
  \ \\
  exec_{2,2}^{(k)}\ \deq\ 
  \begin{conj}
    pc_2^{(k)} = 2\\
    X^{(k)} = Y^{(k)} + 2\ \land\ E_{2,2} = X^{(k)}\\
    pc_2^{(k+1)} = 3\ \land\ pc_1^{(k+1)} = pc_1^{(k)}
  \end{conj}
\end{array}\]

Finally, the overall scheduling constraint is defined as
\[
  sched\ \deq\ round^{(1)}\ \land\ round^{(2)}\ \land\ round^{(3)}
\]
where the formula $round^{(k)}$ representing a single round is defined as
\[\begin{disj}
  pc_1^{(k)} = 3 \land pc_1^{(k+1)} = 3 \land pc_2^{(k)} = 3 \land pc_2^{(k+1)} = 3
  \land Y^{(k+1)} = X^{(k)} \land X^{(k+1)} = X^{(k)}\\
  \begin{conj}
    \begin{disj}
    exec_{1,1}^{(k)} \\
    exec_{1,2}^{(k)} \land E_{1,1} \leq Y^{(k)}\\
    exec_{2,2}^{(k)} \land E_{2,1} \leq Y^{(k)}
    \end{disj}\\
   \begin{disj}
      some\_executable^{(k)} \land Y^{(k+1)} = X^{(k)}\\
      \begin{conj}
        \lnot some\_executable^{(k)}\\
        \begin{disj}
          \begin{conj}
            pc_1^{(k+1)} \neq 3
            \land (pc_2^{(k+1)} \neq 3 \implies E_{prev_1^{(k+1)}} \leq E_{prev_2^{(k+1)}})\\
            Y^{(k+1)} = E_{prev_1^{(k+1)}}
          \end{conj}\\
          \begin{conj}
            pc_2^{(k+1)} \neq 3
            \land (pc_1^{(k+1)} \neq 3 \implies E_{prev_2^{(k+1)}} \leq E_{prev_1^{(k+1)}})\\
            Y^{(k+1)} = E_{prev_2^{(k+1)}}
          \end{conj}
        \end{disj}
      \end{conj}
    \end{disj}
  \end{conj}
\end{disj}\]
and $some\_executable^{(k)}$ is
\[\begin{disj}
  pc_1^{(k+1)} \neq 3 \land (pc_1^{(k+1)}=1 \lor E_{prev_1^{(k+1)}} \leq X^{(k)})\\
  pc_2^{(k+1)} \neq 3 \land (pc_2^{(k+1)}=1 \lor E_{prev_2^{(k+1)}} \leq X^{(k)})
\end{disj}\]

The property required of this program is that thread $2$ executes its non-sleeping statement after all statements of thread $1$, which is expressed as
\[
  \lambda\ \deq\ 
  (pc_1^{(4)} > 2\ \land\ pc_2^{(4)} > 2)\ \implies\ E_{1,2} < E_{2,2}
\]

Yices reports that the formula $sched \land \lnot\lambda$ is unsatisfiable, confirming that the property holds for all executions of the program that respect the timing annotations, as discussed in Section~\ref{sec:moti}. If the annotation of the first statement of thread $1$ is changed to $2$ time units, Yices reports satisfiability, corresponding to a schedule that first executes $l_{1,1}$, then $l_{2,2}$, and finally $l_{1,2}$. In order to simplify experimentation with different values for the timing annotations and sleeping times, our implementation generates symbolic constants for them whose values can easily be changed in the header of the file.

\subsection{Evaluation of Scalability}
\label{sec:eval}

We now present some more experimental results for evaluating how our approach scales. We performed 13 experiments on variations of the producer-consumer example introduced in Section~\ref{sec:moti} that we ran on a PC with a 1.7 GHz Intel Core i7 processor with 8GB memory, using Yices (\emph{version 1.0.39}) as the core engine to perform constraint solving.

In particular, the results in Table \ref{t1} illustrate the scalability of our approach with respect to the number of threads. For these experiments, we consider a pipeline program, which consists of one producer thread $p$, and several copies of consumer threads $c_k$. As shown in the following code snippet, all threads maintain a local variable. The producer thread $p$ initiates and updates the value of its local variable $j_0$. Each consumer thread $c_k$ copies the value of its predecessor's local variable $j_{k-1}$ into its own local variable $j_k$. Our experiments are performed for $k \in \{2,3,5,10,20,100\}$.

\begin{display}
  \begin{tabbing}
    $l_9:$\hspace*{2mm}\= \hspace{3.5cm}\= $l_9:$\hspace*{2mm}\=\kill
    \comment{Thread $p$}      \>\> \comment{Thread $c_{k}$}\\[.3ex]
    $l_1:$ \> \tannot{$1$}        \> $l_3:$ \> sleep($2*k+1$);\\
           \> $j_0 = 0$;          \> $l_4:$ \> \tannot{$2$}\\
    $l_2:$ \> \tannot{$2$}        \>        \> $j_k = j_{k-1}$;\\
           \> $j_0 \mathop{+\!\!=} 2$;\\[.5ex]
    \comment{post-condition:\quad $\displaystyle \bigwedge_{k \geq 1} j_k == j_{k-1}$}
  \end{tabbing}
\end{display}

\begin{table}
\centering
  \begin{tabular}{ l || c | c | c |c }
    \hline
    Test & \#threads & Encoding Time & Execution Time & Conflicts\\ \hline
    1& 2& 0.031s&0.003472s &No\\
    2& 3& 0.031s&0.004341s&No\\
    3& 5&0.033s&0.007635s&No\\
    4&10&0.036s&0.037222s&No\\
    5&20&0.071s&0.3413s&No\\
    6&50&0.585s&6.87579s&No\\
    7&100&3.970s&105.543s&No\\
 
    \hline
  \end{tabular}
  \caption{Scalability in terms of threads}
  \label{t1}
 \end{table}

The experiments in Table~\ref{t2} demonstrate the scalability of our approach when programs have loops that are unrolled to different numbers of iterations. For experiments from $1$ to $5$, we consider a two-threaded producer and consumer program. Each thread has an infinite loop that is unwound $L$ times; $N$ is chosen as $2L+1$, which is big enough to let both threads perform $L$ loop iterations. The post-condition requires that the $i$-th instance of statement $l_5$ is executed between the $i$-th and $(i+1)$-st instances of statement $l_2$. We ran experiments for $L \in \{ 2, 3, 5, 10, 20\}$. 
\bigskip

\begin{display}
  \begin{tabbing}
    $l_9:$\hspace*{2mm}\= \hspace{4.5cm}\= $l_9:$\hspace*{2mm}\=\kill
    \comment{$t_1$: {\it producer}}      \>\> \comment{$t_2$: {\it consumer}}\\[.3ex]
    $l_1:$ \> \tannot{$1$}        \> loop \{\\
           \> $i = 2$;            \> $l_4:$ \> sleep($2$);\\
          loop \{\>  \>    $l_5:$ \> \tannot{$2$}\\   
         $l_2:$ \> \tannot{$2$}  \>  \> $j = i$;\\
          \> $i \mathop{+\!\!=} 2$;  \> \} // $k$ iterations \\
      $l_3:$ 
      \>      sleep($2$);\\
           \} // $k$ iterations\\[.5ex]
    \comment{post-condition:\quad $j == i$ {\it in each round}}
  \end{tabbing}
\end{display}

\begin{table}
 \centering
  \begin{tabular}{ l || c | c | c |c }
    \hline
    Test & $L$ & Encoding Time & Execution Time & Conflicts\\ \hline
    1& 2 & 0.032s&0.00507s &No\\
    2& 3 & 0.033s&0.007074s&No\\
    3& 5 & 0.035s&0.036461s&No\\
    4&10 & 0.036s&1.9266s&No\\
    5&20 & 0.083s&314.761s&No\\
    \hline
     6& 2 & 0.031s&0.017515s&Yes\\
    \hline
  \end{tabular}
 \caption{Scalability in terms of loops}
  \label{t2}
 \end{table}

The experiment $6$ in Table~\ref{t2} is slightly different. The program skeleton is shown below. It is a modified producer-consumer program that consists of two threads. Each thread has one loop, unwound to 10 iterations. Thread $t_1$ updates $a$ using a value $t$, which is chosen randomly. Thread $t_2$ updates $b$ using the value of $a$. The correctness requirement is that the values of $a$ and $b$ are equal at the end of each iteration, that is, the assignment $l_8$ in thread $t_2$ should be executed after $l_2$ and $l_3$ of thread $t_1$ in every iteration.
In this experiment, access conflicts are detected when the program enters into the second iteration of loop. 

\begin{display}
  \begin{tabbing}
    $l_9:$\hspace*{2mm}\= \hspace{4.5cm}\= $l_9:$\hspace*{2mm}\=\kill
    \comment{$t_1$: {\it producer}}      \>\> \comment{$t_2$: {\it consumer}}\\[.3ex]
    $l_1:$ \> \tannot{$1$}        \> $l_6:$ \> \tannot{$1$}\\
           \> $i = 0$;            \>  \> $j=0$;\\
         loop \{\>  \>    $l_7:$ \> sleep($9$);\\   
         $l_2:$ \> \tannot{$2$}  \> loop \{\\
          \> $t= random()$;  \> $l_8:$ \> \tannot{$4$}\\
          $l_3:$ \> \tannot{$5$} \> \> $b=a;$\\ 
      \>      $a= t+2$; \> $l_9:$\> sleep($8$);\\
      $l_4:$\> sleep($10$);\> $l_{10}:$ \> \tannot{$1$}\\
       $l_5:$ \> \tannot{$2$} \> \> $j++;$\\ 
      \>      $i++$; \>\} \\ 
           \} \\[.5ex]  
    \comment{post-condition:\quad $b == a$ {\it in each round}}
  \end{tabbing}
\end{display}

These experiments illustrate the scalability of our approach both in terms of number of threads and number of loops. To push the method to its limit, we intentionally focused on models without conflicts: in this case the solver has to check all possible schedules in order to conclude that the input formula is unsatisfiable. In general, this is more demanding than finding a counter model exhibiting the conflict, as can be seen from the last experiment in Table~\ref{t2}.

\section{Related Work}
\label{sec:dis}

SMT-based constraint solving \cite{nieuwenhuis:solving} has been an active area of research over the last decade, and has led to the existence of many highly efficient tools, such as Yices or Z3 \cite{dutertre:fast,moura:z3}. Due to technological advances and industrial applications, these solutions have attracted much interest and have been applied in different areas, including program analysis and property verification \cite{HS10,BP09,CG12}. Our work proposes an approach to reduce the problem of analyzing multi-threaded programs with time annotations to a constraint solving problem amenable to SMT techniques, which enables us to leverage the strength of existing powerful techniques to solve our problem.

As a fundamental challenge in designing reliable multi-threaded programs, data race detection has attracted significant research efforts. However, existing solutions target the analysis of programs that are based on the use of locks, critical sections and so on \cite{LC02,AE03,PF06,VS11,DR13}, whereas we aim at analyzing multithreaded programs that use time annotations to regulate the program execution and ensure the absence of access conflicts. 

Some recent work studied data race detection using constraint solving. For example, ODR \cite{ASODR} utilizes constraint solving to determine a schedule that satisfies the recorded information. CLAP \cite{HZDCLAP} reproduces concurrency bugs by solving symbolic constraints and monitoring the local execution paths of threads. While that work is similar to ours concerning the use of constraint solving to analyze program execution for finding bugs related to data races or similar concurrency bugs, our problem is different in that it targets the coordination between threads based on time annotations. To the best of our knowledge, our work is the first that analyzes potential conflicts in multi-threaded programs with time annotations using off-the-shelf SMT solvers.

\section{Conclusion}
\label{sec:conclu}

The analysis of timing behavior is fundamental for ensuring the correctness of real-time programs. In particular, multi-threaded real-time programs can achieve synchronization by relying on global time and scheduling constraints. In this paper, we have proposed a representation of such constraints in the language of SMT solvers, and have shown how this encoding can be used to ensure that simple programs satisfy precedence properties. Although the size of the generated formulas is quadratic in the size of the programs, our experiments seem to indicate that modern SMT solvers are powerful enough for this analysis to scale reasonably well. In contrast, we initially experimented with encoding program skeletons as timed automata and using model checking for determining the existence of access conflicts, and this approach did not appear to be scalable.

In this paper, we have only considered programs in which every thread consists of simple straight-line code, possibly within an infinite loop. In particular, we do not handle conditional statements. It is straightforward to extend the approach to non-deterministic choices between alternative branches, and this can be used to abstract from conditional choices by ignoring the values of program variables, and hence from branches that conditional statements would follow in actual executions. For values that can be represented in SMT solvers, such as (bounded or unbounded) integer variables, our approach can be combined with standard SMT-based program analysis techniques~\cite{filliatre:why3,leino:dafny} in order to obtain a more precise verdict, avoiding spurious counter-examples. It will also be interesting to consider scheduling constraints that specify lower and upper bounds, rather than precise running times. Such an analysis will be useful for platforms that cannot guarantee precise execution times corresponding to the timing annotations, but where useful bounds can still be inferred.

Our current approach requires programs that contain timing annotations for ordinary (non-sleeping) statements, as well as sleep statements with fixed sleeping time. An interesting variation would be to provide timing annotations for the ordinary statements, but to leave open the sleeping times. The objective would be to infer sleep statements that ensure certain scheduling constraints, expressed by a formula $\lambda$. This problem can be naturally reduced to integer parameter synthesis for timed automata, a well studied technique, e.g. \cite{ipta}. We intend to explore if this problem can also be solved efficiently using SMT techniques.




\bibliographystyle{eceasst}
\bibliography{bib}

\newcommand{\etalchar}[1]{$^{#1}$}
\begin{thebibliography}{{JSR}13}

\bibitem[AS09]{ASODR}
G.~Altekar, I.~Stoica.
ODR: Output-deterministic Replay for Multicore Debugging.
In Matthews and Anderson (eds.), \emph{ACM SIGOPS Symp. Operating Systems
  Principles}.
SOSP 2009, pp.~193--206.
ACM, Big Sky, Montana, USA, 2009.

\bibitem[BPS09]{BP09}
M.~Botin\v{c}an, M.~Parkinson, W.~Schulte.
Separation Logic Verification of {C} Programs with an {SMT} Solver.
\emph{Electron. Notes Theor. Comput. Sci.} 254:5--23, Oct. 2009.

\bibitem[CG12]{CG12}
A.~Cimatti, A.~Griggio.
Software Model Checking via {IC3}.
In \emph{24th Intl. Conf. Computer Aided Verification (CAV 2012)}.
LNCS~7358, pp.~277--293.
Springer, Berkeley, CA, U.S.A., 2012.

\bibitem[CLL{\etalchar{+}}02]{LC02}
J.-D. Choi, K.~Lee, A.~Loginov, R.~O'Callahan, V.~Sarkar, M.~Sridharan.
Efficient and Precise Datarace Detection for Multithreaded Object-oriented
  Programs.
In \emph{ACM SIGPLAN Conf. Programming Language Design and Implementation}.
PLDI 2002, pp.~258--269.
ACM, Berlin, Germany, 2002.

\bibitem[DM06]{dutertre:fast}
B.~Dutertre, L.~M. de~Moura.
A Fast Linear-Arithmetic Solver for {DPLL(T)}.
In Ball and Jones (eds.), \emph{18th Intl. Conf. Computer-Aided Verification
  (CAV 2006)}.
LNCS~4144, pp.~81--94.
Springer, Seattle, WA, U.S.A., 2006.

\bibitem[EA03]{AE03}
D.~Engler, K.~Ashcraft.
RacerX: Effective, Static Detection of Race Conditions and Deadlocks.
In Scott and Peterson (eds.), \emph{10th ACM Symp. Operating Systems
  Principles}.
SOSP 2003, pp.~237--252.
ACM, Bolton Landing, NY, USA, 2003.

\bibitem[FP13]{filliatre:why3}
J.-C. Filli{\^a}tre, A.~Paskevich.
Why3 - Where Programs Meet Provers.
In Felleisen and Gardner (eds.), \emph{22nd Europ. Symp. Programming (ESOP
  2013)}.
LNCS~7792, pp.~125--128.
Springer, Rome, Italy, 2013.

\bibitem[HSIG10]{HS10}
W.~R. Harris, S.~Sankaranarayanan, F.~Ivan\v{c}i\'{c}, A.~Gupta.
Program Analysis via Satisfiability Modulo Path Programs.
In Hermenegildo and Palsberg (eds.), \emph{37th Ann. ACM SIGPLAN-SIGACT Symp.
  Principles of Programming Languages}.
POPL 2010, pp.~71--82.
ACM, Madrid, Spain, 2010.

\bibitem[HZD13]{HZDCLAP}
J.~Huang, C.~Zhang, J.~Dolby.
CLAP: recording local executions to reproduce concurrency failures.
In Boehm and Flanagan (eds.), \emph{34th ACM SIGPLAN Conf. Programming Language
  Design and Implementation}.
PLDI 2013, pp.~141--152.
ACM, Seattle, WA, U.S.A., 2013.

\bibitem[JLR13]{ipta}
A.~Jovanovic, D.~Lime, O.~H. Roux.
Integer Parameter Synthesis for Timed Automata.
In \emph{19th Intl. Conf. Tools and Algorithms for the Construction and
  Analysis of Systems (TACAS 2013)}.
LNCS~7795, pp.~401--415.
Springer, Rome, Italy, 2013.

\bibitem[{JSR}13]{jsr302}
{JSR 302 Expert Group}.
JSR 302: Safety Critical Java Technology.
Java Specification Requests, 2013.
\url{https://jcp.org/en/jsr/detail?id=302}.

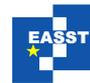
\bibitem[Lam94]{lamport:howtowrite}
L.~Lamport.
How to Write a Long Formula.
\emph{Formal Aspects of Computing} 6(5):580--584, 1994.

\bibitem[Lei13]{leino:dafny}
K.~R.~M. Leino.
Developing verified programs with {Dafny}.
In Notkin et~al. (eds.), \emph{35th Intl. Conf. Software Engineering}.
ICSE 2013, pp.~1488--1490.
ACM, San Francisco, CA, U.S.A., 2013.

\bibitem[MB08]{moura:z3}
L.~M. de~Moura, N.~Bj{\o}rner.
Z3: An Efficient {SMT} Solver.
In Ramakrishnan and Rehof (eds.), \emph{14th Intl. Conf. Tools and Algorithms
  for the Construction and Analysis of Systems (TACAS 2008)}.
LNCS~4963, pp.~337--340.
Springer, Budapest, Hungary, 2008.

\bibitem[NOT06]{nieuwenhuis:solving}
R.~Nieuwenhuis, A.~Oliveras, C.~Tinelli.
Solving SAT and SAT Modulo Theories: From an abstract
  Davis--Putnam--Logemann--Loveland procedure to DPLL(T).
\emph{J. ACM} 53(6):937--977, 2006.

\bibitem[PFH06]{PF06}
P.~Pratikakis, J.~S. Foster, M.~Hicks.
LOCKSMITH: Context-sensitive Correlation Analysis for Race Detection.
In \emph{ACM SIGPLAN Conf. Programming Language Design and Implementation}.
PLDI 2006, pp.~320--331.
ACM, Ottawa, Ontario, Canada, 2006.

\bibitem[RD13]{DR13}
C.~Radoi, D.~Dig.
Practical Static Race Detection for Java Parallel Loops.
In \emph{Intl. Symp. Software Testing and Analysis}.
ISSTA 2013, pp.~178--190.
ACM, Lugano, Switzerland, 2013.

\bibitem[SPPH10]{schoeberl:worst-case}
M.~Schoeberl, W.~Puffitsch, R.~U. Pedersen, B.~Huber.
Worst-case execution time analysis for a Java processor.
\emph{Software Practice and Experience} 40(6):507--542, 2010.

\bibitem[SVEH11]{VS11}
T.~Sheng, N.~Vachharajani, S.~Eranian, R.~Hundt.
RACEZ: A Lightweight and Non-Invasive Race Detection Tool for Production
  Applications.
In Taylor et~al. (eds.), \emph{33rd Intl. Conf. Software Engineering}.
ICSE 2011, pp.~401--410.
ACM, Honolulu, HI, USA, 2011.

\bibitem[TW04]{TWRealTime04}
L.~Thiele, R.~Wilhelm.
Design for Timing Predictability.
\emph{Journal Real-Time Systems.} 28(2-3):157--177, Nov. 2004.

\end{thebibliography}

\end{document}